\documentclass{osa-article}
\journal{osajournal}
\articletype{Research Article}

\usepackage{hyperref}
\usepackage{mathtools, mathrsfs}
\usepackage{caption}
\usepackage{subcaption}
\usepackage{color}

\begin{document}

\title{Illumination Pattern Design with Deep Learning for Single-Shot Fourier Ptychographic Microscopy}

\author{Yi Fei Cheng,\authormark{1,$^\dagger$} Megan Strachan,\authormark{1,$^\dagger$} Zachary Weiss,\authormark{1,$^\dagger$} Moniher Deb,\authormark{2} Dawn Carone,\authormark{2} Vidya Ganapati\authormark{1,*}}

\address{\authormark{1}Swarthmore College, Engineering Department, 500 College Ave, Swarthmore, PA 19081} 
\address{\authormark{2}Swarthmore College, Biology Department, 500 College Ave, Swarthmore, PA 19081} 

\address{\authormark{$^\dagger$}Equal contribution.}
\email{\authormark{*}vganapa1@swarthmore.edu} 

\homepage{ganapati.swarthmore.edu}

\begin{abstract}
Fourier ptychographic microscopy allows for the collection of images with a high space-bandwidth product at the cost of temporal resolution. In Fourier ptychographic microscopy, the light source of a conventional widefield microscope is replaced with a light-emitting diode (LED) matrix, and multiple images are collected with different LED illumination patterns. From these images, a higher-resolution image can be computationally reconstructed without sacrificing field-of-view. We use deep learning to achieve single-shot imaging without sacrificing the space-bandwidth product, reducing the acquisition time in Fourier ptychographic microscopy by a factor of 69. In our deep learning approach, a training dataset of high-resolution images is used to jointly optimize a single LED illumination pattern with the parameters of a reconstruction algorithm. Our work paves the way for high-throughput imaging in biological studies.
\end{abstract}

\section{Introduction}

The space-bandwidth product collected by a microscope is proportional to the product of the field-of-view and resolution \cite{goodman_introduction_2005}. High field-of-view and high resolution is desirable; however, the space-bandwidth product is limited by the number of pixels in the image sensor. Additionally, optical aberrations limit the creation of a high numerical aperture and low-magnification objective \cite{zheng_wide-field_2013}. For a standard widefield microscope, we must trade off field-of-view and resolution in a single image. To increase the space-bandwidth product, we can take multiple images, such as by mechanically scanning a sample with a high numerical aperture objective. 

In the last five years, Fourier ptychographic microscopy \cite{zheng_wide-field_2013} has emerged as a alternative to mechanical scanning to achieve a high space-bandwidth product. In Fourier ptychographic microscopy, the illumination source of a standard widefield microscope is replaced by a matrix of light-emitting diodes (LEDs). Multiple images of the sample are captured with different patterns of LED illumination. The collected image stack is then used for computational reconstruction of the phase and amplitude of the sample, with spatial frequencies exceeding that of the objective's numerical aperture. In each illumination pattern, different spatial frequencies are modulated into the pass-band of the optical system. Though the resolution of each collected image is determined by the microscope objective's numerical aperture, the final reconstructed complex object has a resolution corresponding to a numerical aperture that is the sum of the illumination and the objective numerical apertures. In the computational reconstruction, the field-of-view of the lower-resolution collected images is not sacrificed, leading to an enhanced space-bandwidth product. The advantages of Fourier ptychographic microscopy over slide scanning are the elimination of mechanical scanning, which may lead to higher temporal resolution, and greater depth-of-field from the use of a lower numerical aperture objective \cite{zheng_wide-field_2013}. 

Originally, Fourier ptychographic microscopy involved the sequential illumination and image capture of each LED in the matrix \cite{zheng_wide-field_2013}. Subsequent work showed that acquisition time could be reduced by using multiplexed patterns of LEDs  \cite{tian_multiplexed_2014, tian_computational_2015} or an adaptive illumination strategy \cite{zhang_self-learning_2015}. Though the approach in \cite{tian_computational_2015} demonstrates sub-second image stack acquisition for Fourier ptychographic reconstruction, sophisticated hardware control is required. In this work, we aim to obtain Fourier ptychographic reconstruction with a single image from a single LED illumination pattern. In order to do so, we trade off the generality of the Fourier ptychographic method. In standard Fourier ptychographic microscopy, we generally don't make assumptions about the sample we are imaging. Here, by assuming that our sample comes from a particular probability distribution, we aim to reduce the measurements we need down to a single image, greatly improving the temporal resolution without sacrificing space-bandwidth product.

Due to loss of phase information in the captured intensity images, the Fourier ptychographic object reconstruction is generally performed with iterative algorithms, which can be computationally intensive. Recently, it has been shown that these iterative algorithms can be replaced with deep learning \cite{kappeler_ptychnet_2017, nguyen_deep_2018, boominathan_phase_2018}. The parameters of a deep neural network are trained using examples of input low-resolution image stacks and output high-resolution objects. Though this training process is computationally intensive, once trained, reconstruction can be on the order of tens of seconds \cite{nguyen_deep_2018}.

For iterative reconstruction in Fourier ptychographic microscopy, the number of collected images needed for successful reconstruction is determined by the overlap of the images in Fourier space \cite{dong_sparsely_2014}. In the deep learning approach, due to the use of prior information regarding sample distribution in developing the reconstruction algorithm, the number of collected images needed is reduced \cite{kappeler_ptychnet_2017, nguyen_deep_2018, boominathan_phase_2018}.
 
In \cite{robey_optimal_2018}, we proposed the joint optimization of a single LED illumination pattern with the parameters of a post-processing algorithm to achieve a high space-bandwidth-time product in Fourier ptychographic microscopy. The joint optimization took the form of training a deep convolutional neural network. We demonstrated single-shot Fourier ptychographic microscopy for different simulated sample distributions, retraining for each sample distribution. Our work showed that the joint optimization caused the mutual information between the single collected low-resolution image and the high-resolution object to increase.

Data-driven joint optimization of hardware parameters with the post-processing algorithm for imaging tasks has been previously performed in \cite{chakrabarti_learning_2016, horstmeyer_convolutional_2017, elmalem_learned_2018, haim_depth_2018, hershko_multicolor_2018, diederich_using_2018, kellman_physics-based_2018}. In \cite{horstmeyer_convolutional_2017, kellman_physics-based_2018}, the LED illumination patterns in Fourier ptychographic microscopy are optimized. In \cite{horstmeyer_convolutional_2017}, two LED illumination patterns are co-optimized with a neural network to classify biological cells infected with malaria. In \cite{kellman_physics-based_2018}, 2-4 LED illumination patterns are co-optimized with an iterative algorithm for complex object reconstruction under the weak object approximation.

In this work, we experimentally validate the approach in \cite{robey_optimal_2018}, demonstrating single-shot Fourier ptychographic microscopy for a given sample type. Our results with a single image replicate with high fidelity the results from an iterative approach with a 69 image stack. Our demonstration of single-shot, high space-bandwidth product paves the way for high-throughput biological studies.

\section{Fourier Ptychographic Microscopy}

\subsection{Theory}

\begin{figure}
\centering
\begin{subfigure}[b]{.47\textwidth}
  \centering
  \includegraphics[width=\textwidth]{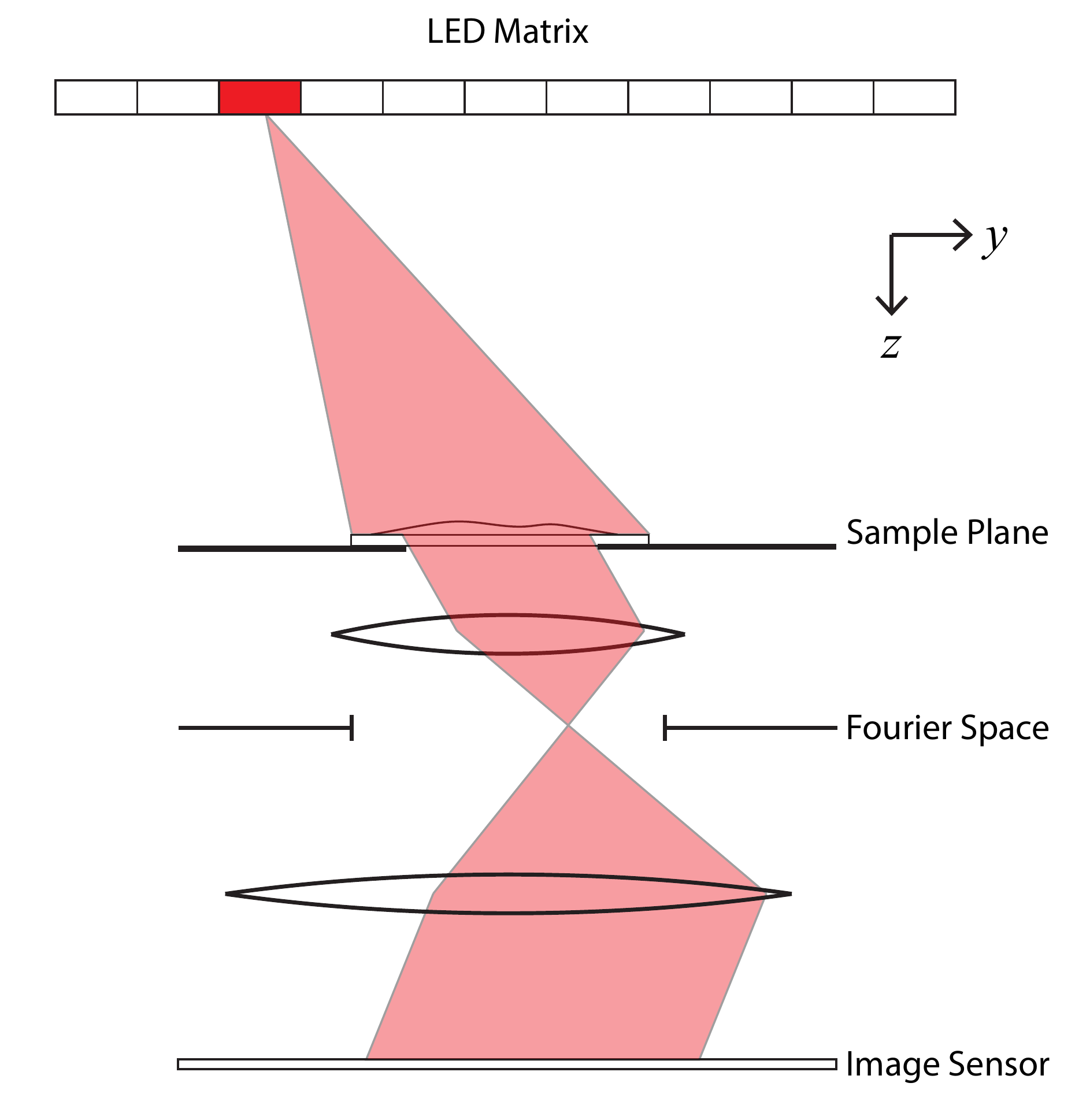}
  \caption{Schematic of the Fourier ptychographic microscopy setup. An LED matrix replaces the illumination source of a conventional microscope.}
  \label{fig:schematic}
\end{subfigure}\hfill%
\begin{subfigure}[b]{.47\textwidth}
  \centering
  \includegraphics[width=\textwidth]{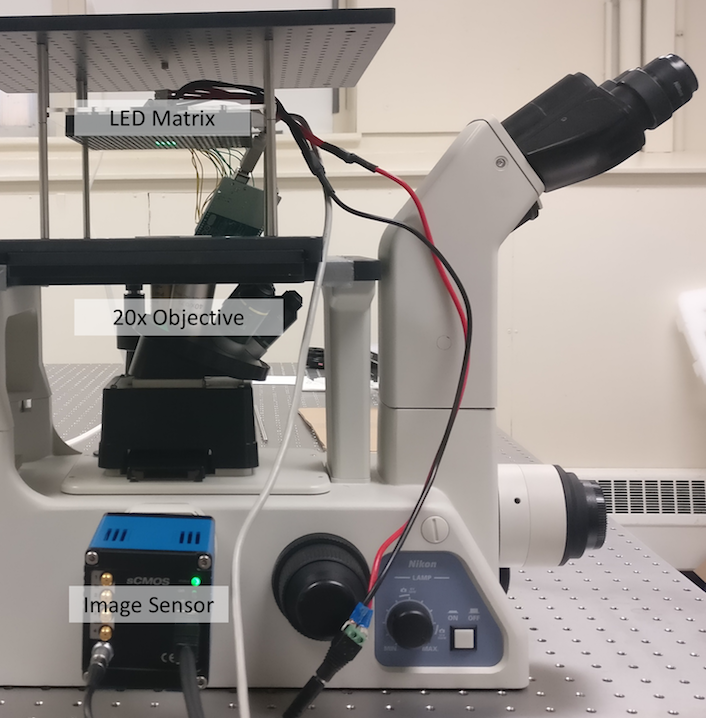}
  \caption{Experimental Fourier ptychographic microscopy setup. We utilize a 20$\times$ magnification objective with a numerical aperture of 0.5 in our setup.}
  \label{fig:graph}
\end{subfigure}
\caption{}
\label{fig:FPMdiagrams}
\end{figure}

In Fourier ptychographic microscopy, the light source in a conventional widefield microscope is replaced by an LED matrix. We assume that the LED matrix is sufficiently far enough away that the light entering the sample from a single LED can be approximated as a plane wave. Fig.~\ref{fig:FPMdiagrams}(a) shows a sample illuminated by a single LED. The sample scatters the entering plane wave, and the exiting wave is imaged through a microscope objective with some magnification and numerical aperture (NA).

The spatial profile of the plane wave from the LED entering the sample is described by $e^{i 2 \pi \vec{u_l} \cdot \vec{r}}$ where $ \vec{u_l} \cdot \hat{x} = u_{l,x} $ is the spatial frequency in the $x$-direction and $ \vec{u_l} \cdot \hat{y} = u_{l,y} $ is the spatial frequency in the $y$-direction. The magnitude of $\vec{u_l}$ is $\mid\vec{u_l}\mid = \frac{1}{\lambda}$, where $\lambda$ is the wavelength of the light. 

We use the thin transparency approximation to describe the sample as a complex transmission function $o(x,y)$. The field immediately after the sample can be described as the object transmission function multiplied by the illumination plane wave at $z=0$:
\begin{equation} 
o(x,y) e^{ i 2 \pi \left(u_{l,x} x + u_{l,y} y \right)}.
\label{eq:exit-wave}
\end{equation}

If the 2-dimensional Fourier transform of $o(x,y)$ is given as $O(u_x, u_y)$, the field in Eqn.~\ref{eq:exit-wave} in Fourier space is simply a shift, $O(u_x - u_{l,x}, u_y - u_{l,y})$. The field is multiplied in Fourier space by the pupil function of the objective, $P(u_x, u_y)$, a low-pass filter. In the case of no aberrations, $P(u_x, u_y)$ is unity within a circle with radius $\frac{\text{NA}}{\lambda}$ and zero elsewhere. The image sensor records the intensity $I$ of this field in real space:

\begin{equation}
I = \mid \mathscr{F}^{-1} \{ P(u_x, u_y)O(u_x - u_{l,x}, u_y - u_{l,y}) \} \mid^2 ,
\label{eq:lowres}
\end{equation}

\noindent where $\mathscr{F}^{-1}$ denotes the inverse 2-dimensional Fourier transform \cite{tian_multiplexed_2014, yeh_experimental_2015}. Due to the low-pass filtering operation, the intensity image contains information from only a portion of spatial frequencies of the original object.

The previous discussion assumes illumination from a single LED, where it is assumed that the LED emits a coherent wave. In a matrix of LEDs, with multiple LEDs illuminated, we assume that each LED emits a coherent wave that is mutually incoherent with the waves emitted from the other LEDs. The intensity image recorded from multiplexed LED illumination is:

\begin{equation}
I = \sum \limits_{l=1}^n c_{l} \mid \mathscr{F} \{ P(u_x, u_y)O(u_x - u_{l,x}, u_y - u_{l,y}) \} \mid^2 ,
\end{equation}

\noindent where $n$ is the total number of illuminated LEDs and $c_l$ is the intensity of LED $l$ \cite{tian_multiplexed_2014}.

\subsection{Experimental Setup}

Our experimental setup is shown in Fig.~\ref{fig:FPMdiagrams}(b). A programmable matrix of 32 by 32 LEDs with 4 mm pitch (Adafruit) is mounted on an inverted microscope (Nikon Eclipse TE300). In this work, we utilize only the 69 centermost LEDs of the 32 by 32 matrix. The 69 centermost LEDs correspond to brightfield illumination. The distance $z$ from the LED matrix to the sample plane is 69.5 mm and the wavelength of light from the LED array is approximately 518 nm. The microscope objective is 20$\times$ magnification with a numerical aperture of 0.5 (Nikon CFI Plan Fluor). The image sensor captures 16-bit images with 2048 by 2048 pixels and a pixel size of 6.5 $\mu$m (pco.edge 4.2 LT). For iterative Fourier ptychographic microscopy, we sequentially collect an image stack of 69 images. Each image has single LED illumination and an exposure of 2 seconds. The use of the 69 centermost LEDs results in a synthetic numerical aperture of 0.75 for the Fourier ptychographic microscope.

\subsection{Iterative Object Reconstruction}

We implement the iterative reconstruction algorithm as a computational graph with the Python package TensorFlow \cite{abadi_tensorflow_2016}. TensorFlow allows for easy implementation of gradient-based optimization procedures through the use of automatic differentiation. Our approach to using automatic differentiation for the computational reconstruction is similar to that in \cite{ghosh_adp_2018, jiang_solving_2018}. It should be emphasized that in this section, we describe Tensorflow being used as a basis for iterative optimization, not for data-driven machine learning. 

We start with an initial guess of the high-resolution object, $o(x,y)$:

\begin{equation}
o(x,y) = \sqrt{ \frac{1}{69}\sum \limits_{l = 0}^{68} I_l(x,y)}, 
\end{equation}

\noindent where $I_l(x,y)$ is the low-resolution image collected with LED $l$ illuminated. We subtract the background from every $I_l(x,y)$ as in \cite{tian_multiplexed_2014}.

From our guess of $o(x,y)$, we computationally simulate the low-resolution image stack using Eqn.~\ref{eq:lowres}, obtaining $I_l^{g}(x,y)$, the simulated low-resolution image for LED $l$. We evaluate how close $o(x,y)$ is to the actual object by computing

\begin{equation}
L = \sum \limits_{l = 0}^{68} \left( \sqrt{I_l(x,y)} - \sqrt{I_l^{g}(x,y)} \right)^2,
\label{eq:loss}
\end{equation}
\noindent

\noindent where $L$ is the loss function. We compare amplitudes instead of intensities in the loss function for better tolerance to noise and model mismatch \cite{yeh_experimental_2015}.

The gradients $L$ with respect to the real and imaginary parts of each pixel of $o(x,y)$ are calculated through automatic differentiation in TensorFlow. The guess of the object $o(x,y)$ is updated using the Adam optimization algorithm \cite{kingma_adam_2014} with these gradients. We also calculate the gradients of $L$ with respect to the phase at each pixel of the pupil function $P(u_x, u_y)$ and update the phase of $P(u_x, u_y)$ as in \cite{ou_embedded_2014, tian_multiplexed_2014}.

To parallelize across multiple computers, we split our collected image stacks into 16 patches with some overlap and perform this iterative optimization procedure on each patch. At the end, we merge the reconstructed $o(x,y)$ patches, using linear interpolation in the overlapping regions. Our training rate is $2 \times 10^{-1}$ and we run the Adam optimizer for 3,000 iterations on each patch. 

\begin{figure}[htbp]
    \centering
    \includegraphics[scale=0.4]{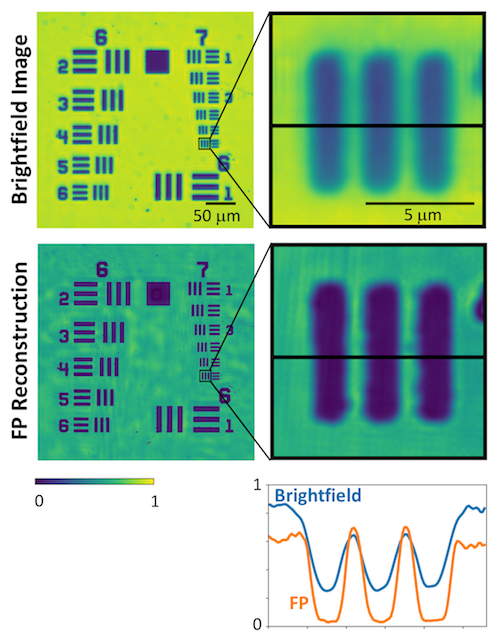}
    \caption{Brightfield image intensity (top) and intensity of the iterative Fourier ptychographic reconstruction (bottom) of a target slide. We see improved image quality in the iterative Fourier ptychographic reconstruction. The zoomed in region has a resolution of 228.0 line pairs per millimeter.}
    \label{fig:target}
\end{figure}

Fig.~\ref{fig:target} shows the zoomed-in reconstruction of a target slide (Thorlabs R1DS1P). We compare the Fourier ptychographic (FP) reconstruction intensity, $\mid{o(x,y)}\mid^2$, with the brightfield intensity, ${ \sum \limits_{l = 0}^{68} I_l(x,y)}$. The brightfield intensity emulates the image that would be captured with all the LEDs simultaneously illuminated. The displayed Fourier ptychographic reconstruction and brightfield intensity images are divided by their average values and normalized.  We see in Fig.~\ref{fig:target} that our Fourier ptychographic microscopy procedure yields improved image quality over standard brightfield microscopy. 

\section{Deep Learning}

\subsection{Overview}

\begin{figure}[htbp]
    \centering
    \includegraphics[scale=0.4]{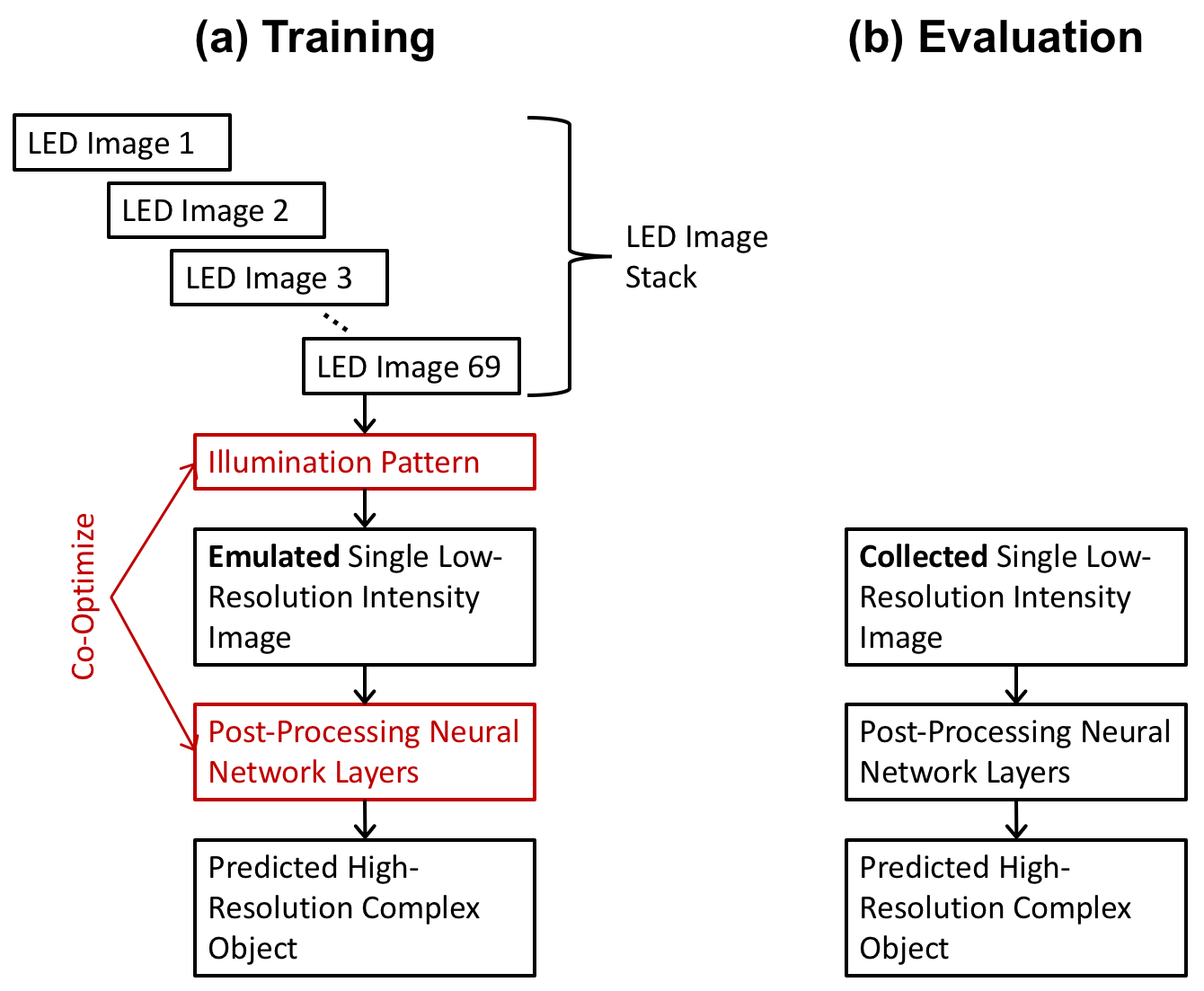}
    \caption{Overview of the training and evaluation steps in our data-driven joint hardware and software optimization.}
    \label{fig:diagram}
\end{figure}

A high-level overview of our general deep learning framework in shown in Fig.~\ref{fig:diagram}. In this work, there are two main phases: training and evaluation.

In the training phase, our aim is to find a single LED illumination pattern and a post-processing algorithm for high-resolution object reconstruction. In this phase, we first use iterative Fourier ptychographic microscopy methods to construct a training dataset. We collect a stack of low-resolution images for each sample, with each image corresponding to illumination from a single LED. From the low-resolution image stack, we iteratively reconstruct the high-resolution complex object. For each sample in our dataset, we have a low-resolution image stack and corresponding high-resolution complex object. 

Using the training dataset, we train the deep neural network shown diagrammatically in Fig.~\ref{fig:diagram}(a). In the deep neural network, the LED illumination pattern is represented by a trainable variable. For each example, we input the low-resolution image stack and computationally emulate the single low-resolution intensity image we would expect given the LED illumination pattern. The single low-resolution image is then fed to the post-processing neural network layers. The post-processing layers are in the form of a convolutional neural network \cite{lecun_object_1999} and are parametrized by trainable weights and biases. The final output of the network is the prediction of the high-resolution complex object. We optimize the trainable variables of the network in order to minimize the difference between the predicted high-resolution complex object and the actual object from the iterative reconstruction using the entire low-resolution image stack.

Once the training phase is complete, we have an optimized LED illumination pattern and a non-iterative reconstruction algorithm. In the evaluation phase, shown in Fig.~\ref{fig:diagram}(b), we first  program our LED matrix on the Fourier ptychographic microscope with this static pattern. Then, for each sample, we take a single image and process it through our trained post-processing neural network layers to get the high-resolution complex object. Image acquisition and post-processing are fast in the evaluation phase, enabling high-throughput or real-time live imaging.

\subsection{Training}

For the training dataset, we collect low resolution image stacks with sequential LED scanning. For each field-of-view imaged, we collect 5 stacks and use the average value in the iterative Fourier ptychographic reconstruction of the complex object. We collected 20 fields-of-view over 2 sample slides for the training dataset in this work. The samples used in this work are Tig-1 normal human fibroblast cells. Tig-1 cells grown in monolayer on glass coverslips were fixed with 4\% paraformaldeyhyde and treated with 2.5\% Trypsin prior to staining with Giemsa stain in Gurr's Buffer (KaryoMAX, ThermoFisher). Coverslips were mounted onto glass slides and sealed with CytoSeal (Richard Allan Scientific, VWR) prior to imaging. 

We represent the neural network diagrammed in Fig.~\ref{fig:diagram}(a) as a computational graph in TensorFlow. The structure of the computational graph is given in \cite{robey_optimal_2018}. The post-processing layers are in the form of a convolutional neural network with residual layers, inspired by the neural network architecture in \cite{mao_image_2016}. 

We initialize the illumination pattern by giving each LED a brightness value uniformly distributed from 0 (i.e. the LED is not illuminated) to 1 (i.e. the LED is at its maximum brightness). The illumination pattern is also parametrized by the exposure time, which we initialize to 200 milliseconds. The exposure is constrained to be from 0 to 2000 milliseconds and each LED brightness value is constrained between 0 and 1. Each image in the input low-resolution stack is collected with a single LED at maximum brightness and maximum exposure. We can emulate the single low-resolution image from the illumination pattern as:

\begin{equation}
I_s(x,y) = \epsilon \sum  \limits_{l = 0}^{68} c_l I_l(x,y)
\end{equation}
 
\noindent where $I_s(x,y)$ is the simulated low-resolution image, $\epsilon$ is the exposure normalized by 2000 milliseconds, and $c_l$ is the brightness value for LED $l$. Both $c_l$ and $\epsilon$ are variable parameters that are trained during the optimization procedure.
 
Our image sensor collects a 16-bit image, so we truncate $I_s(x,y)$ at a minimum value of 0 and a maximum value of $2^{16} - 1$. We approximate the effect of quantization by adding uniform noise from 0 to 1 on each pixel of $I_s(x,y)$, redrawing the value of the random variable every evaluation of the TensorFlow computational graph. We also add a Gaussian approximation of Poisson noise to $I_s(x,y)$. Every pixel of $I_s(x,y)$ is processed by:

\begin{equation}
\text{max} \left( \frac{\sqrt{I_{low} \times m} \times g + I_{low} \times m}{m}, 0 \right),
\label{eq:noise}
\end{equation}

\noindent where $m$ is a multiplicative factor chosen to fit the Poisson noise of our experimental setup and $g$ is drawn from a normal random distribution. A higher $m$ corresponds to higher signal-to-noise ratio. Again, we redraw the value of $g$ at every evaluation of the TensorFlow computational graph.

\begin{figure}[htbp]
    \centering
    \includegraphics[scale=0.4]{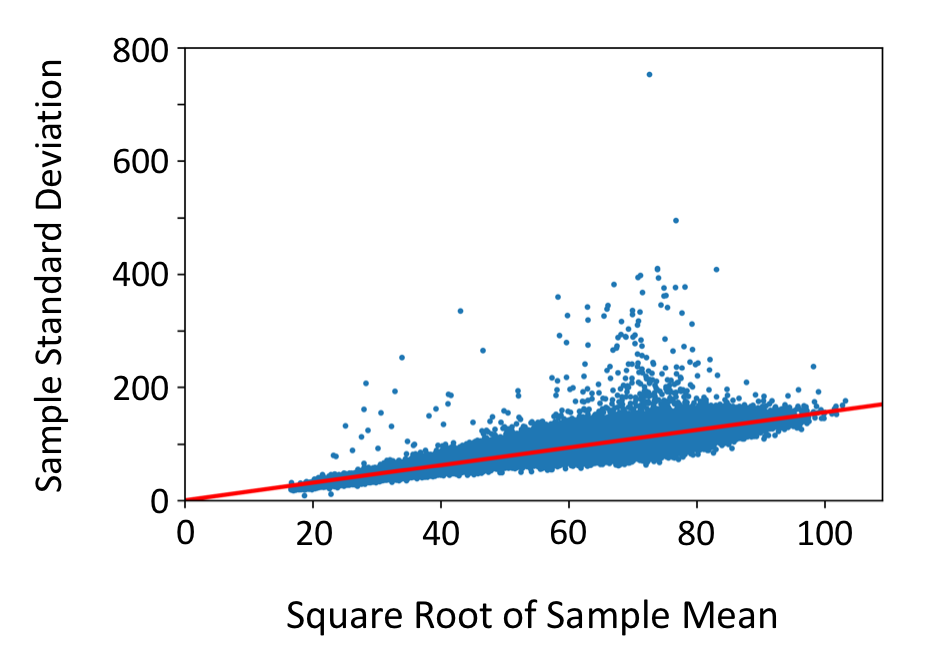}
    \caption{We include a Gaussian approximation of Poisson noise in our computational model of the Fourier ptychographic microscope. To estimate $m$ in Eqn.~\ref{eq:noise}, we took 100 images with the center LED illuminated. This plot shows the square root of the sample mean vs. the sample standard deviation at every point of the image stack. The slope $s = \sqrt{\frac{1}{m}}$.}
    \label{fig:pnm}
\end{figure}

To find $m$, the multiplicative factor for Poisson noise, we took 100 images of a sample with the center LED illuminated. We plotted the square root of the sample mean vs. the sample standard deviation at every point of the image stack, see Fig.~\ref{fig:pnm}. We found a linear fit between the two quantities with slope $s=0.41$, and $m = \frac{1}{s^2}$.

The noisy $I_s(x,y)$ is then inputted into the post-processing neural network layers which output a prediction of the high-resolution complex object. We minimize the objective function described in \cite{robey_optimal_2018} to optimize all the trainable parameters of our computational graph. The objective function includes the mean-squared difference between the actual and predicted complex object and their first gradients \cite{robey_optimal_2018}.

\subsection{Fine-Tuning}

To correct any inaccuracies in our emulation of the low-resolution image $I_s(x,y)$, such as in our noise model, we perform a fine-tuning step. In the fine-tuning step, we first collect the low-resolution image stack as before, but for each field-of-view, we also collect an image with the optimized LED illumination pattern. For this work, we collected data from 30 regions over 3 sample slides.

On each region, we then perform iterative Fourier ptychographic reconstruction as before. Thus, we get a dataset of optimized LED pattern images and corresponding reconstructed complex objects. 

With this dataset, we fine-tune the parameters of the post-processing layers. We initialize with the parameters found in the initial training step and continue training by feeding the optimized LED pattern image directly into the post-processing layers. As before, in the initial training step, we optimize the variable parameters in order to minimize our objective function. However, we do not update the LED illumination pattern parameters in this step.

\subsection{Evaluation}

Once training and fine-tuning are complete, we collect a single image with the optimized LED illumination pattern. We then feed the collected image into the post-processing neural network layers to non-iteratively reconstruct the high-resolution complex object. For comparison, we collect the full single LED illumination low-resolution stack and perform iterative Fourier ptychographic reconstruction. To show that the neural network is not overfitting, the evaluation data is taken from a new sample slide not used in training or fine-tuning.

\section{Results and Discussion}

\begin{figure}[htbp]
    \centering
    \includegraphics[scale=0.35]{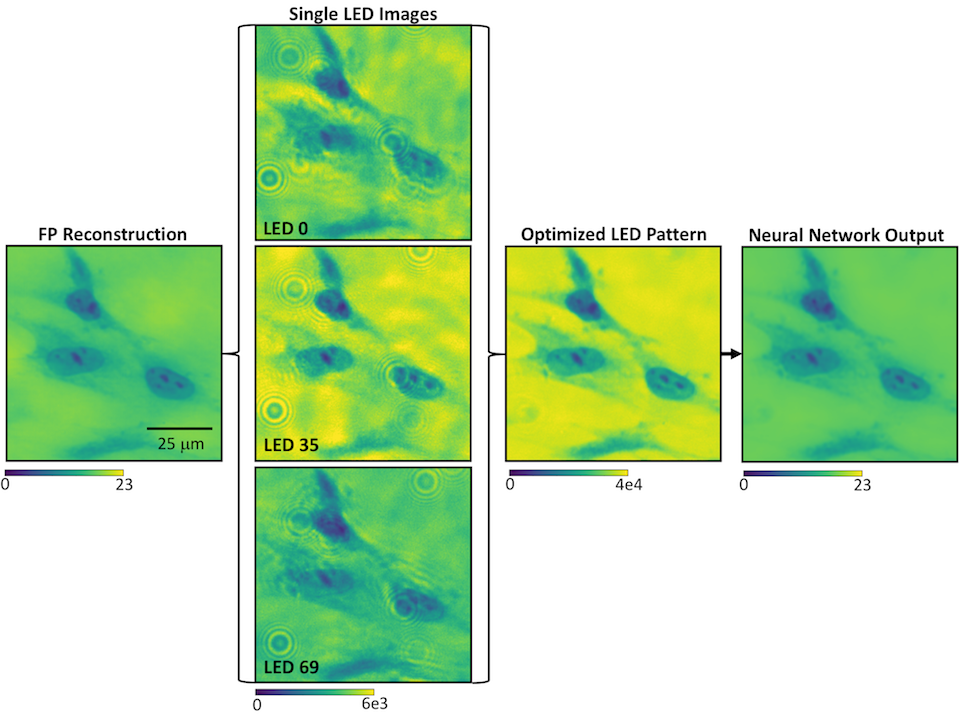}
    \caption{An image patch from the evaluation dataset. From left to right: the amplitude of the iterative Fourier ptychographic reconstruction, 3 images from the 69 low-resolution image stack each with single LED illumination, the collected image with the optimized LED illumination pattern, and the amplitude of the neural network reconstruction. In the 3 images from the low-resolution stack, there is a circular artifact corresponding to the image of the single source LED.}
    \label{fig:pipeline}
\end{figure}

In Fig.~\ref{fig:pipeline}, we show an image patch from the evaluation data. The collected optimized LED pattern image is approximately a linear combination of the 69 single LED illumination images. The optimized LED pattern image is inputted into the trained neural network post-processing layers, and the amplitude of the neural network output is shown. For comparison, the amplitude of the iterative Fourier ptychographic reconstruction is also shown. 

\begin{figure}[htbp]
    \centering
    \includegraphics[scale=0.35]{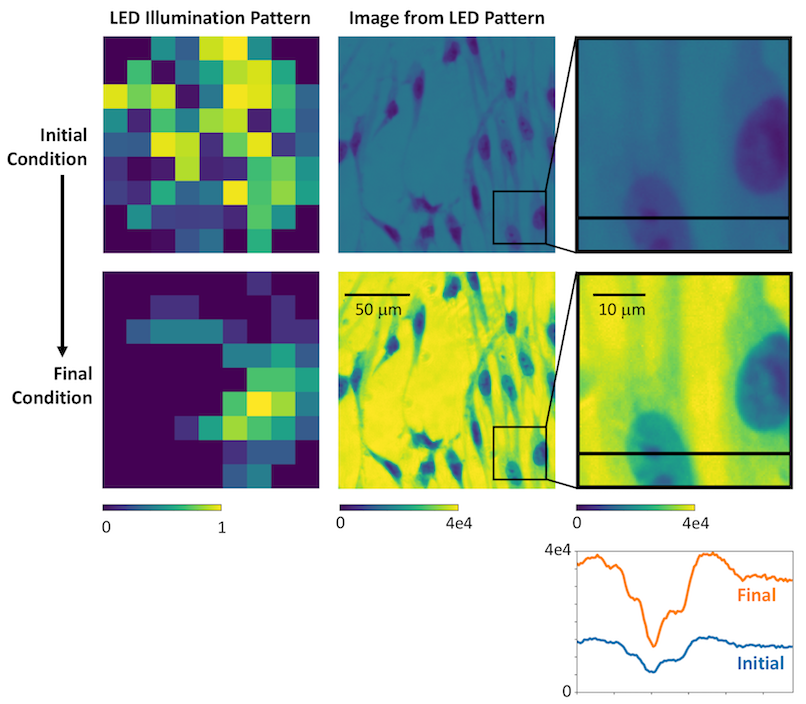}
    \caption{Initial and final LED illumination patterns with the corresponding emulated image patches. The exposures for the initial and final images are 200 and 1140 milliseconds, respectively. We visually see an increase in contrast and information content in the final image with the optimized LED illumination pattern.}
    \label{fig:if}
\end{figure}

In \cite{robey_optimal_2018}, we obtained better performance by updating the LED illumination pattern during training than by leaving the LED illumination pattern fixed at a randomly chosen or uniform illumination pattern. We showed in \cite{robey_optimal_2018} that with our joint optimization procedure, the mutual information increased between the low-resolution collected image and the high-resolution complex object. We collect more information with the optimized illumination than with a randomly selected illumination pattern. Fig.~\ref{fig:if} shows the initial and final optimized LED illumination pattern, along with corresponding emulated image patches. We can visually see the contrast increase with the final optimized LED illumination pattern; it appears that the final image has more information content. The image patches shown are taken from the evaluation dataset; the analysis in Fig.~\ref{fig:if} was done after training and fine-tuning were complete as to not bias the training procedure.

\begin{figure}
\centering
\begin{subfigure}[b]{.5\textwidth}
  \centering
  \includegraphics[width=\textwidth]{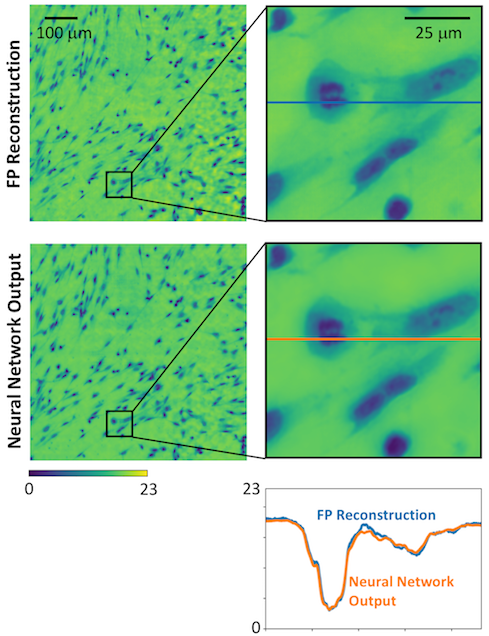}
  \caption{}
  \label{fig:ex1}
\end{subfigure}\hfill%
\begin{subfigure}[b]{.5\textwidth}
  \centering
  \includegraphics[width=\textwidth]{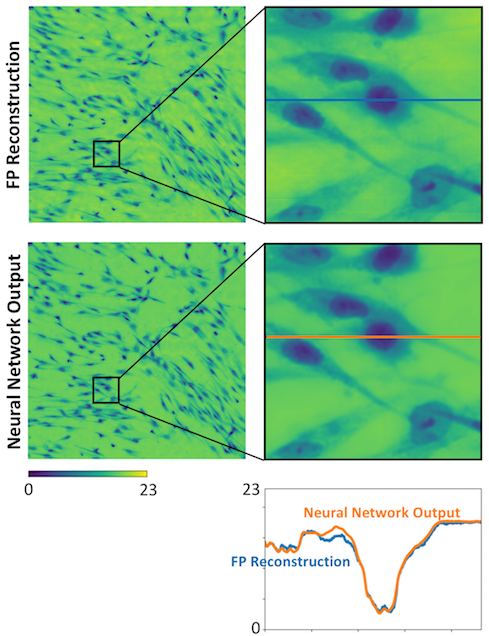}
  \caption{}
  \label{fig:ex2}
\end{subfigure}\hfill%
\begin{subfigure}[b]{.5\textwidth}
  \centering
  \includegraphics[width=\textwidth]{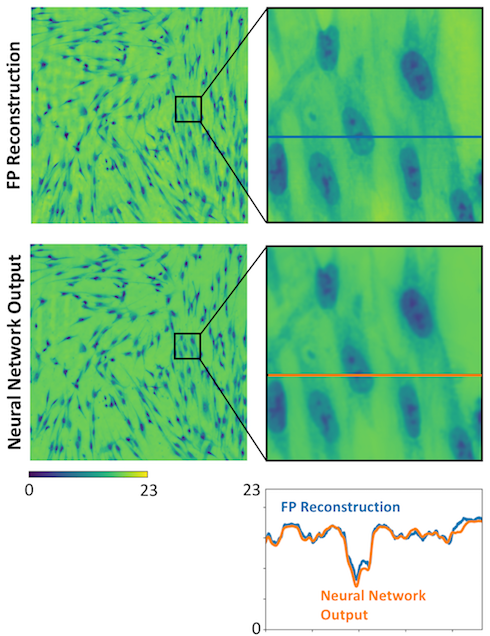}
  \caption{}
  \label{fig:ex3}
\end{subfigure}
\caption{Examples from the evaluation dataset, comparing amplitude of the iterative Fourier ptychographic reconstruction (the ground truth) and the neural network output. We see high fidelity between the two reconstructions. The neural network output requires only 1 collected image, representing a factor of 69 decrease in acquisition time. Linescans cut through cells in different phases of the cell cycle; cells are in (a) anaphase, (b) prophase, and (c) interphase.}
\label{fig:examples}
\end{figure}

Fig.~\ref{fig:examples} shows 3 examples from the evaluation dataset. We show the full collected field-of-view in each example, as well as a zoomed in image patch and linescan. We compare the amplitude of the iterative Fourier ptychographic reconstruction with the amplitude of the neural network reconstruction. The iterative Fourier ptychographic reconstruction requires the full 69 low-resolution image stack, while the neural network reconstruction only requires a single collected image. We achieve high fidelity between both reconstructions. The neural network reconstruction appears to have filtered out noise in the iterative Fourier ptychographic reconstruction. We hypothesize that since the noise varies randomly in each training example, the neural network was not able to learn how to reconstruct the noise, effectively filtering it out.

\section{Conclusion}

We have developed and demonstrated a data-driven framework for imaging. By using past knowledge, we are able to improve the speed of computational imaging modalities. In this case, we achieved single-shot Fourier ptychographic microscopy without loss of space-bandwidth product, a speed-up of 69 times. The initial overhead in collecting the training dataset and training the computational graph enables future fast image collection and object reconstruction. The framework presented allows for probing dynamic processes, performing high-throughput studies, and reducing data storage needs.

This work demonstrates a high space-bandwidth-time product for imaging Tig-1 cells. Each new sample distribution requires retraining to determine new hardware and software parameters for image collection and reconstruction, respectively. However, using the techniques of transfer learning may reduce the training time and dataset size needed for new sample distributions \cite{pan_survey_2010}.

\section*{Acknowledgments}

We would like to thank Carr Everbach, Ed Jaoudi, and Cassy Burnett for assistance with material acquisition; Jeffrey Knerr, Douglas Campbell, and Alexander Robey for technical support; Russell Prigodich for material fabrication; and the Surdna Foundation (Y. F. C.) and the Swarthmore College Summer Research Fellowship (M. S. and Z. W.) for funding.

\bibliography{FPM_2}

\end{document}